\pdfoutput=1
\documentclass[11pt,a4paper]{article}
\usepackage{graphicx} 
\usepackage{dcolumn,booktabs}
\usepackage{bm} 
\usepackage{amssymb}
\usepackage[hmargin=2.cm,vmargin=3.5cm]{geometry}

\newcommand{\op}{\ensuremath{\mathcal{O}}}

\newcommand{\be}{\begin{equation}} 
\newcommand{\ee}{\end{equation}} 
\newcommand{\bea}{\begin{eqnarray}}  
\newcommand{\eea}{\end{eqnarray}}
\newcommand{\bs}{\begin{split}} 
\newcommand{\es}{\end{split}}

\newcommand{\units}[1]{~\mathrm{#1}}

\newcommand{\SM}{\mathrm{SM}}

\begin{document}

\begin{center}

{\Large {\bf Global Constraints on Lepton-Quark Contact Interactions}}\\

\vspace*{1cm}
{\bf Jorge de Blas$^a$}, {\bf Mikael Chala$^b$} and {\bf Jos\'e Santiago$^b$}
\vspace{0.5cm}

$^a$Department of Physics, University of Notre Dame,\\
Notre Dame, IN 46556, USA

\vspace{.5cm}

$^b$CAFPE and Departamento de F\'{\i}sica Te\'orica y del Cosmos,\\
Universidad de Granada, E-18071 Granada, Spain

\end{center}

\begin{abstract}
The Large Hadron Collider can do precision physics at a
level that is competitive with electroweak
precision constraints when probing physics beyond the Standard
Model. We present a simple yet general
parameterization of the effect of an arbitrary number of lepton-quark
contact interactions on any di-lepton observable at hadron colliders. This
parameterization can be easily adopted by the experimental
collaborations to put bounds on arbitrary combinations of lepton-quark
contact interactions. We compute the corresponding bounds from current di-lepton
resonance searches at the LHC and find that they are competitive with and often
complementary to indirect constraints from electroweak precision
data. We combine all current constraints in a global analysis to obtain the most
stringent bounds on lepton-quark contact interactions. 
We also show that the high-energy
phase of the LHC has a unique potential in terms of discovery and
discrimination power among different types of lepton-quark contact
interactions.
\end{abstract} 

\renewcommand{\thefootnote}{\arabic{footnote}}
\setcounter{footnote}{0}

\date{\today}

\vspace{1 cm}

The Large Hadron Collider (LHC) is a discovery machine at the energy
frontier. However, it can also do precision physics, probing physics
beyond the Standard Model (SM) in a complementary and even competitive way
to electroweak precision data (EWPD). 
Model-independent bounds
on departures from the SM can be systematically computed by means of
effective Lagrangians. Assuming the SM particle content and symmetries
and neglecting lepton number violation, the leading corrections arise
from dimension-six operators
\begin{equation}
\mathcal{L}_{\mathrm{eff}}=\mathcal{L}_{\mathrm{SM}} +\sum_i \frac{\alpha_i}{
\Lambda^2} \mathcal{O}_i + \ldots,
\end{equation}
where $\mathcal{L}_{\mathrm{SM}}$ is the SM Lagrangian and
$\Lambda$ is the cut-off scale at which the effective Lagrangian
ceases to be valid. 
The list of required
operators was systematically classified for the first time by Ref.~\cite{Buchmuller:1985jz}. Several redundancies were discussed by many authors~\cite{Rattazzi:1988ye} and the first list of truly independent operators
was given in Ref.~\cite{Grzadkowski:2010es} (see Ref.~\cite{Contino:2013kra} for a recent alternative).
In this work  
we consider the constraints that current
searches for di-lepton resonances imply on lepton-quark four-fermion
interactions (see Refs.~\cite{DiBartolomeo:1997mb,Carpentier:2010ue} for related analyses).
The most general lepton-quark four-fermion interaction can be
parameterized in terms of the following dimension-six operators:
\begin{eqnarray}
\hspace{0.08cm}\op_{lq}^{(1)}\hspace{-0.08cm}&=&(\bar{l}\gamma^\mu l) (\bar{q}\gamma_\mu
q),
\quad
\op_{lq}^{(3)}=(\bar{l}\sigma_I \gamma^\mu l)
(\bar{q}\sigma_I\gamma_\mu q),
\nonumber \\[0.cm]
\op_{eu}&=&(\bar{e}\gamma^\mu e) (\bar{u}\gamma_\mu u),
\quad
\hspace{-0.15cm}\op_{ed}\hspace{0.07cm}=(\bar{e}\gamma^\mu e) (\bar{d}\gamma_\mu d),
\nonumber \\[0.cm]
\hspace{-0.025cm}\op_{lu}\hspace{0.025cm}&=&(\bar{l} \gamma^\mu l) (\bar{u} \gamma_\mu u),
\quad
\hspace{-0.04cm}\op_{ld}\hspace{0.11cm}=(\bar{l} \gamma^\mu l) (\bar{d} \gamma_\mu d),
\label{4F_5_6}\\[0.cm]
\hspace{-0.cm}\op_{qe}\hspace{0.cm}&=&(\bar{q} \gamma^\mu q) (\bar{e} \gamma_\mu e),
\quad
\hspace{-0.08cm}\op_{qde}\hspace{-0.06cm}=(\bar{l} e) (\bar{d} q),
\nonumber \\[0.cm]
\hspace{0.05cm}\op_{lq\epsilon}\hspace{-0.05cm}&=&(\bar{l} e)\epsilon (\bar{q}^T u),
\quad
\hspace{0.33cm}\op_{ql\epsilon}\hspace{0.02cm}=(\bar{q} e) \epsilon (\bar{l}^T u),
\nonumber
\end{eqnarray}
where $l$ and $q$ are the SM lepton and quark doublets; $e$, $u$
and $d$ denote the SM lepton and quark singlets; $\sigma_I$ are the Pauli matrices; 
and $\epsilon=\mathrm{i}\sigma_2$. Flavor indices are
not explicitly shown. 

In the following we will assume that the operator
coefficients are flavor-diagonal and family-universal in the quark sector. This guarantees the absence of contributions
to quark flavor changing neutral currents (FCNC) from the first seven operators ($\op_{lq}^{(1)}$ - $\op_{qe}$). Note that, while EWPD is sensitive to
the couplings to all families, the largest constraints from LHC searches come from the coupling to valence quarks. Therefore, we will only consider the contributions to LHC observables
from the couplings to the first-generation quarks.
Regarding the lepton sector, we
consider three different options: a flavor-diagonal family-universal coupling; interactions aligned with the SM fermion basis in a way that there are couplings only to electrons; and couplings only to muons.

For the last three operators ($\op_{qde}$, $\op_{lq\epsilon}$, $\op_{ql\epsilon}$) the above assumptions still result on minimal flavor violating interactions in the quark sector. (Coefficients proportional to the quark mixing matrices would be required to remove FCNC.) Even in the absence of FCNC, these three operators still give sizable contributions to charged-current interactions mediating rare decays. Such contributions are strongly constrained, e.g., the experimental value of $R_\pi=\Gamma(\pi^+\rightarrow \nu e^+)/\Gamma(\pi^+\rightarrow \nu \mu^+)$ sets bounds $|\alpha_i|/\Lambda^2\lesssim {\cal O}(10^{-3}) \mbox{ TeV}^{-2}$ for corrections only to the muon channel, and two orders of magnitude smaller for the electron channel \cite{Carpentier:2010ue}. As we will see, such limits are significantly stronger than the LHC and EWPD constraints discussed here. Therefore, we will not include these three operators in our numerical analyses, although they are included in all our equations for the sake of completeness.

The contribution of the operators in (\ref{4F_5_6}) to Drell-Yan production reads, at
the partonic level,
\begin{eqnarray}
48 \pi \frac{d\sigma}{d\hat{t}}(\bar{u}u\to \ell^+ \ell^-)&=&
\left[ 
\left|
\mathcal{A}^{\mathrm{SM}}_{u_L \ell_R}
+
\frac{\alpha_{qe}}{\Lambda^2}\right|^2
+
\left|
\mathcal{A}^{\mathrm{SM}}_{u_R \ell_L}
\!+
\frac{\alpha_{lu}}{\Lambda^2}\right|^2+\frac{1}{2\Lambda^4}
\Big[|\alpha_{ql\epsilon}|^2
+\mathrm{Re}(\alpha_{lq\epsilon}\alpha^\ast_{ql\epsilon})\Big]\right] 
\!\frac{\hat{t}^2}{\hat{s}^2}
\nonumber \\
&+&
\left[ 
\left|
\mathcal{A}^{\mathrm{SM}}_{u_L \ell_L}
+\frac{\alpha_{lq}^{(1)}-\alpha_{lq}^{(3)}}{\Lambda^2}\right|^2
+
\left|
\mathcal{A}^{\mathrm{SM}}_{u_R \ell_R}
+
\frac{\alpha_{eu}}{\Lambda^2}\right|^2-\frac{1}{2\Lambda^4}
\mathrm{Re}(\alpha_{lq\epsilon}\alpha^\ast_{ql\epsilon})\right]
\frac{\hat{u}^2}{\hat{s}^2} \nonumber\\
&+&
\frac{1}{2\Lambda^4}
\Big[|\alpha_{lq\epsilon}|^2
+\mathrm{Re}(\alpha_{lq\epsilon}\alpha^\ast_{ql\epsilon})\Big]
,
\label{sigma:uu}
\\
48 \pi \frac{d\sigma}{d\hat{t}}(\bar{d}d\to \ell^+ \ell^-)&=&
\left[ 
\left|
\mathcal{A}^{\mathrm{SM}}_{d_L \ell_R}
+
\frac{\alpha_{qe}}{\Lambda^2}\right|^2
+
\left|
\mathcal{A}^{\mathrm{SM}}_{d_R \ell_L}
+
\frac{\alpha_{ld}}{\Lambda^2}\right|^2\right] 
\frac{\hat{t}^2}{\hat{s}^2}
\nonumber\\
&+&
\left[ 
\left|
\mathcal{A}^{\mathrm{SM}}_{d_L \ell_L}
+
\frac{\alpha_{lq}^{(1)}+\alpha_{lq}^{(3)}}{\Lambda^2}\right|^2
+
\left|
\mathcal{A}^{\mathrm{SM}}_{d_R \ell_R}
+
\frac{\alpha_{ed}}{\Lambda^2}\right|^2\right] 
\frac{\hat{u}^2}{\hat{s}^2} 
+\frac{|\alpha_{qde}|^2}{2\Lambda^4}
,
\nonumber
\end{eqnarray}
where we have defined 
\[
\mathcal{A}^{\mathrm{SM}}_{\psi \phi} =\frac{e^2 Q_\psi Q_\phi}{\hat{s}} +
\frac{g_{\psi}g_{\phi}}{\hat{s}-m_Z^2+\mathrm{i} m_Z \Gamma_Z},
\]
with 
$g_\psi = \frac{g}{c_W}[T^3_\psi - s_W^2 Q_\psi]$, $Q$ the electric charge, $T^3$ the third component of weak isospin, $g$ the $SU(2)_L$ coupling, $m_Z$ and
$\Gamma_Z$ the $Z$-boson mass and width, and
$s_W (c_W)$ the sine (cosine) of the weak angle.
This result completes previous partial calculations \cite{DiBartolomeo:1997mb,Eichten:1983hw}.

Eqs.~(\ref{sigma:uu})
allow us to parameterize any
di-lepton observable at the LHC in the presence of arbitrary
lepton-quark contact interactions. 
Any such observable can be written in terms of the cross section in a
particular region of phase space as measured by experiments,
$\sigma$, which can in turn be written in the form of a master
equation with a small number of parameters. In the limit of large
di-lepton invariant masses, the master equation takes the following
simple form
\begin{eqnarray}
\sigma &=&
\sigma^{\SM} 
+\frac{1}{\Lambda^2} \sum_{q=u,d}\left[
F_1^q A_1^q  +F_2^q A_2^q 
\right]
+
\frac{1}{\Lambda^4} \sum_{q=u,d} \left[
G_1^q B^q_1 +G_2^q B^q_2+G_3^q B^q_3
\right],  \label{master:formula}
\end{eqnarray}
where we have neglected corrections proportional to $m_Z^2/\hat{s} \ll 1$. 
The coefficients $A^{u,d}_{1,2}$ and
$B^{u,d}_{1,2,3}$ encode the dependence on the four-fermion operators
\begin{eqnarray}
A_1^u & = & 
[e^2 Q_u Q_e + g_{u_L} g_{e_L}] (\alpha^{(1)}_{lq} -\alpha^{(3)}_{lq})
+[e^2 Q_u Q_e + g_{u_R} g_{e_R}] \alpha_{eu},
\nonumber 
\\
A_2^u & = & 
[e^2 Q_u Q_e + g_{u_L} g_{e_R}] \alpha_{qe}
+[e^2 Q_u Q_e + g_{u_R} g_{e_L}] \alpha_{lu},
\nonumber 
\\
A_1^d & = & 
[e^2 Q_d Q_e + g_{d_L} g_{e_L}] (\alpha^{(1)}_{lq} +\alpha^{(3)}_{lq})+[e^2 Q_d Q_e + g_{d_R} g_{e_R}] \alpha_{ed},
\nonumber 
\label{AB:coefficients}
\\
A_2^d & = & 
[e^2 Q_d Q_e + g_{d_L} g_{e_R}] \alpha_{qe}
+[e^2 Q_d Q_e + g_{d_R} g_{e_L}] \alpha_{ld},
\nonumber
\\
B^u_1&=&
4(\alpha_{lq}^{(1)}-\alpha_{lq}^{(3)})^2+4\alpha_{eu}^2
-2 \mathrm{Re}(\alpha_{lq\epsilon}  \alpha_{ql\epsilon}^\ast),
\\
B^u_2&=&
4\alpha_{qe}^2+4\alpha_{lu}^2+2|\alpha_{ql\epsilon}|^2
+2 \mathrm{Re}(\alpha_{lq\epsilon}  \alpha_{ql\epsilon}^\ast),
\nonumber \\
B^u_3&=&
2|\alpha_{lq\epsilon}|^2
+2 \mathrm{Re}(\alpha_{lq\epsilon}  \alpha_{ql\epsilon}^\ast),
\nonumber \\
B^d_1&=&
4(\alpha_{lq}^{(1)}+\alpha_{lq}^{(3)})^2+4\alpha_{ed}^2,
\nonumber \\
B^d_2&=&
4\alpha_{qe}^2+4\alpha_{ld}^2,
\nonumber \\
B^d_3&=&
2|\alpha_{qde}|^2.
\nonumber 
\end{eqnarray}
$\sigma^{\SM}$, $F^{u,d}_{1,2}$ and $G^{u,d}_{1,2,3}$ on the other
hand depend on the particular phase space region we are considering
for the observable we want to compute, and 
encode the effects of the parton distribution functions and the cuts involved in the
experimental analyses. 
A further simplification can be obtained in the case of
forward-backward symmetric observables, for which we can impose the
following extra conditions:
\begin{equation}
\left . \begin{array}{l}
F^{u}_1=F^u_2,\quad 
F^d_1=F^d_2\\
G^{u}_1=G^u_2,\quad 
G^d_1=G^d_2
\end{array}\right\}
\mbox{ (symmetric observables)},
\label{symmetric:condition}
\end{equation}
since the corresponding contributions are related by a
$\hat{t}\leftrightarrow \hat{u}$ exchange. Finally, for symmetric 
observables for which the
experimental acceptance is approximately constant along the detector coverage,
we can also impose
\begin{equation}
G^u_3=3 G^u_1, \quad
G^d_3=3 G^d_1, \label{svsut}
\end{equation}
relating the contributions proportional to $\hat{s}$ with those
proportional to $\hat{t}$ and $\hat{u}$. 

This master 
equation, Eqs.~(\ref{master:formula}) and (\ref{AB:coefficients}), constitutes the
main result of the present paper. It can be easily
adopted by the LHC experimental collaborations and, once they have
computed the specific values of the observable-dependent coefficients
and their uncertainties, 
bounds on arbitrary combinations of lepton-quark contact
interactions can be easily obtained.

To show how this can be done, we have 
computed the observable-dependent coefficients by
implementing the effective operators in 
\texttt{FeynRules} 1.6~\cite{Christensen:2008py}. We have then used
\texttt{MadGraph 5}~\cite{Alwall:2011uj} to generate di-lepton events
at the partonic level, 
\texttt{Pythia} 6~\cite{Sjostrand:2006za} for hadronization and showering
and \texttt{Delphes} 3.0.9~\cite{Ovyn:2009tx} for fast detector simulation.
We have implemented the latest ATLAS~\cite{ATLAS} and CMS~\cite{CMS} di-lepton
searches using all the collected luminosity at the LHC with $\sqrt{s}=8$ TeV.
Di-electron and di-muon final states are studied separately in
these analyses. 
We have 
considered results for the following bins in the di-lepton invariant
mass (in TeV) to avoid contamination from non-Drell Yan backgrounds
\begin{eqnarray}
\textrm{CMS}&:& b_1=[0.9,1.3],~b_2=[1.3,1.8],~b_3=[1.8,-], \nonumber \\
\textrm{ATLAS}&:& b_4=[1.2,3], \label{bin:definition}
\end{eqnarray}
resulting in a total of 8 bins, counting electrons and muons.
\begin{table}
\centering
\begin{tabular}{ccccccccc}
\toprule
 & $b_1 (e)$  & $b_2 (e)$  & $b_3 (e)$  & $b_4 (e)$ 
 & $b_1 (\mu)$  & $b_2 (\mu)$  & $b_3 (\mu)$  & $b_4 (\mu)$ 
\\
\hline
$N_{\SM}$ & 32.6 & 4.68 & 0.60 & 8.72 & 37.0 & 5.38 & 0.74 & 9.44 \\
$F^u_1$ & 2514 & 731 & 202 & 1324 & 2746 &  811 & 251 & 1410 \\
$F^d_1$ & 1484 & 359 & 80.2 & 677 & 1590 & 481 & 93.6 & 775 \\
$G^u_1$ & 346 & 203 & 116 & 404 & 376 & 219 & 134 & 415 \\
$G^d_1$ & 200 & 106 & 46.1 & 199 & 219 & 118 & 53.0 & 207 \\
\hline
$N_{\mathrm{Obs}}$ & 41 & 4 & 0 & 10 & 49 & 11 & 1 & 8 \\
\bottomrule
\end{tabular}
\caption{Observable-dependent coefficients for di-lepton LHC searches
(see Eq.~(\ref{bin:definition}) for details). The observable considered is the 
number of events on each bin. The coefficients $F_1^{u,d}$ and $G_1^{u,d}$ are
in units of TeV$^{-2}$ and TeV$^{-4}$, respectively.
Eqs.~(\ref{symmetric:condition}) and (\ref{svsut}) should be used to
fix the remaining parameters. The observed number of
events is also 
reported in each case.\label{master:parameters:nevs8}}
\end{table}
The observable considered is the number of events on each bin. This is
a symmetric observable for which the experimental acceptances are reasonably
constant along the detector coverage. Thus we can use the simplifying
conditions Eqs.~(\ref{symmetric:condition}) and (\ref{svsut}). 
We have checked that these simplifying conditions lead to the correct
number of events within a $3\%$ uncertainty.
The corresponding parameters for our master equation are reported in
Table~\ref{master:parameters:nevs8}, together with the actual number
of observed events. Note that only a small number of simulations are
required to obtain these parameters. This is an important advantage
for experimental collaborations that rely on expensive full detector
simulations. 
In particular they would have to compute the expected
number of events in the SM (separately for signal and background) plus
the expected number of events in the presence of just two operators,
that can be taken for instance $\mathcal{O}_{eu}$ and
$\mathcal{O}_{ed}$, for two values of the corresponding
coefficients. This would suffice to fully generate our master equation
for an arbitrary combination of lepton-quark contact interactions.
We have 
 implemented all the
operators and extensively tested the validity of our master equation
and the approximations in Eqs.~(\ref{symmetric:condition}) and (\ref{svsut}).

Once we have the prediction for the number of events in each bin and
experiment for arbitrary combinations of quark-lepton contact
interactions we can obtain the corresponding bounds on the
coefficients of such operators. For the sake of concreteness,
we consider in the following that only one operator is present at a
time. We use the $\mathrm{CL}_s$ method~\cite{Beringer:1900zz} to
obtain the $95\%$ confidence level (C.L.) bound on the coefficients of the different
operators. The
combination of different bins is performed by defining an effective
$\chi^2$ function for each bin,
\begin{equation}
 \chi^2 = 2 [\mathrm{Erf}^{-1}(1-\mathrm{CL}_s)]^2,
 \label{EffChi2}
\end{equation}
and adding the $\chi^2$ of all bins. This allows a direct
combination with constraints from EWPD and is exact in the limit of
a large number of events. Given the fact that we have several bins
with a small number of events, we have tested the validity of such an
approximation by computing the probability density function of the combined $\mathrm{CL}_s$
test statistic by Monte Carlo simulation using the \texttt{TLimits}
\texttt{Root} class. The resulting bounds agree with our approximation
within $10\%$ on average, but they depart by about $20\%$ in some
cases. To account for this uncertainty, we have imposed
a $20\%$ penalty on the coefficients of the effective operators when
computing the LHC $\chi^2$. This means that the \textit{coefficients} of the
different effective operators are multiplied by a 0.8 factor before
inserting them in the calculation of the LHC $\chi^2$. 
With this penalty the bounds obtained with
the $\chi^2$ are always conservative as compared with the exact bounds
computed with the Monte Carlo method. We present the corresponding
bounds in columns $2-4$ of Table \ref{LHC_EW_Lim} for
the three different flavor options mentioned in the introduction.
The limits range from $0.02$ to $0.13$ TeV$^{-2}$. This also applies
to the last three operators in (\ref{4F_5_6}), and thus justifies neglecting
them in our analysis, since the $R_\pi$ constraints make their effects invisible
at the LHC at $\sqrt{s}=8$ TeV.

\begin{table}[t]
\centering
{\small
\begin{tabular}{c ccc ccc}
\toprule
&\multicolumn{6}{c}{$95\%$ C.L. limits on
$\frac{\alpha_i}{\Lambda^2}~\left[\mathrm{TeV}^{-2}\right]$}\\[+0.05cm] 
&\multicolumn{3}{c}{LHC $(\sqrt{s}=8~\mathrm{TeV},~{\cal
L}\sim 20\units{fb}^{-1})$}&\multicolumn{3}{c}{Electroweak precision
data}\\[+0.05cm] 
${\cal O}_i$&Universal&Only $e$&Only $\mu$&Universal&Only $e$&Only $\mu$\\
\hline
${\cal O}_{lq}^{(1)}$&$\left[-0.032,0.073\right]$&$\left[-0.040,0.082\right]$&$\left[-0.043,0.084\right]$&
$\left[-0.012,0.055\right]$&$\left[-0.012,0.055\right]$&$\left[-0.620,0.669\right]$\\[+0.05cm]
${\cal O}_{lq}^{(3)}$&$\left[-0.106,0.019\right]$&$\left[-0.118,0.026\right]$  &$\left[-0.126,0.026\right]$&
$\left[-0.006,0.012\right]$&$\left[-0.006,0.012\right]$&$\left[-0.169,0.694\right]$\\[+0.05cm]
${\cal O}_{eu}$ &$\left[-0.032,0.102\right]$&$\left[-0.042,0.113\right]$  &$\left[-0.044,0.117\right]$&
$\left[-0.097,0.017\right]$&$\left[-0.097,0.017\right]$&$-$\\[+0.05cm]
${\cal O}_{ed}$&$\left[-0.107,0.068\right]$&$\left[-0.123,0.084\right]$  &$\left[-0.128,0.086\right]$&
$\left[-0.077,0.040\right]$&$\left[-0.077,0.040\right]$&$-$\\[+0.05cm]
${\cal O}_{lu}$ &$\left[-0.043,0.079\right]$&$\left[-0.054,0.090\right]$  &$\left[-0.056,0.093\right]$&
$\left[-0.041,0.095\right]$&$\left[-0.045,0.092\right]$&$\left[-0.335,0.889\right]$\\[+0.05cm]
${\cal O}_{ld}$&$\left[-0.096,0.076\right]$&$\left[-0.112,0.093\right]$  &$\left[-0.117,0.095\right]$&
$\left[-0.021,0.106\right]$&$\left[-0.020,0.107\right]$&$\left[-1.337,1.407\right]$\\[+0.05cm]
${\cal O}_{qe}$&$\left[-0.040,0.058\right]$&$\left[-0.049,0.068\right]$  &$\left[-0.051,0.070\right]$&
$\left[-0.055,0.011\right]$&$\left[-0.055,0.011\right]$&$-$\\
\bottomrule
\end{tabular}
}
\caption{Comparison of the different $95\%$ C.L. limits on lepton-quark contact interactions. The three different flavor realizations
discussed in the introduction are denoted ``Universal'', ``Only $e$'' and ``Only $\mu$'', respectively. 
In all cases we assume diagonal and family-universal interactions with quarks. A dash (``$-$'') is used to indicate those cases
where the data cannot bound the corresponding operator.\label{LHC_EW_Lim}} 
\end{table}

Lepton-quark contact interactions
also contribute to precision observables, and therefore are
indirectly constrained by EWPD. 
These limits are dominated by low-energy measurements (e.g., atomic parity
violation experiments) and by the $e^+e^-\rightarrow \mathrm{hadrons}$ data
taken at energies above the $Z$ pole at LEP2. The electroweak bounds
for all the dimension-six interactions that can be generated at
tree level and can interfere with the SM were computed
in Refs.~\cite{Han:2004az,delAguila:2011zs}.
The EWPD fits in this work include all the
updates discussed in the analysis of the electroweak constraints
in Ref.~\cite{deBlas:2012qp}, the latest values of $\alpha_S$ and the top mass, and the
final results of $e^+e^-\rightarrow \bar{f}f$ at LEP2 \cite{Schael:2013ita}. 
In all cases, we assume real values for the dimension-six operator coefficients.
We extend the results in Ref.~\cite{delAguila:2011zs} for the first seven operators in
(\ref{4F_5_6}), for the case of diagonal and universal quarks interactions, 
and the different lepton flavor hypotheses discussed in the introduction.\footnote{In 
Ref.~\cite{delAguila:2011zs} all fermion interactions
are assumed to be diagonal and family-universal. Also, a
different basis for four-fermion interactions is employed. In particular, the operators $\op_{lu,ld,qe}$ in Ref.~\cite{delAguila:2011zs}
are related to those in this paper by the Fierz
reordering $(\overline{\psi_L^1}\gamma^\mu \psi_L^2)(\overline{\xi_R^3}\gamma_\mu\xi_R^4)=-2~\!(\overline{\psi_L^1}\xi_R^4)
(\overline{\xi_R^3}\psi_L^2).$}
The corresponding bounds are shown in the last three
columns of Table \ref{LHC_EW_Lim}.

Several conclusions can be extracted from the results in
Table \ref{LHC_EW_Lim}. Even though indirect constraints from EWPD
are still in many cases more stringent than those from LHC
searches, the latter are already quite competitive in general 
and in some cases much superior.
The operators involving muons are very poorly constrained by EWPD.
In these cases the LHC constraints are more
than an order of magnitude more stringent. 
The previous most stringent bounds on these operators come
 from Tevatron data \cite{TevConstr}. They are discussed in Ref.~\cite{Carpentier:2010ue}
 and are weaker than the ones from LHC data.
Also there is quite often (see, e.g., $\op_{eu,lu,ld,qe}$) a nice
complementarity between both 
results, with each experimental data set (EWPD
or LHC) improving the worst limit derived from the other. 
It should be noted, however, that, due to the different energies probed
by each set, the range of validity of the effective description is
different in each case. In particular, since we are probing energies
up to $\sim 3$ TeV in LHC searches, the bounds we have
obtained are only valid if the coefficients of the effective operators
satisfy
\begin{equation}
\alpha \gtrsim 9 \frac{\alpha}{\Lambda^2}\Big|_{\mathrm{max}}
\mbox{ or }
\alpha \lesssim 9 \frac{\alpha}{\Lambda^2}\Big|_{\mathrm{min}},
\end{equation}
where $\alpha/\Lambda^2|_{\mathrm{max(min)}}$ stands for the upper
bound in the case of a positive (negative) $\alpha$.

Once we have obtained the bounds on lepton-quark contact interactions
from LHC searches and EWPD, we can consider the
bounds obtained from a joint analysis of both data sets. 
This global analysis provides the most stringent constraints
over the lepton-quark contact interactions under consideration.
The combination is performed by adding the
effective $\chi^2$ in Eq.~(\ref{EffChi2}) for all bins from all the
LHC searches to the $\chi^2$ of the
electroweak fit. The $95\%$ C.L. bounds we have obtained with this procedure
are reported in Table \ref{EW_Comb_Lim}.

\begin{table}
\centering
\begin{tabular}{c c c c}
\toprule
&\multicolumn{3}{c}{$95\%$ C.L. Combined limits on $\frac{\alpha_i}{\Lambda^2}~\left[\mathrm{TeV}^{-2}\right]$}\\[+0.05cm]
${\cal O}_i$& Universal & Only $e$ & Only $\mu$ \\
\hline
${\cal O}_{lq}^{(1)}$&$\left[-0.011,0.053\right]$&$\left[-0.012,0.053\right]$  &$\left[-0.042,0.084\right]$ \\[+0.05cm]
${\cal O}_{lq}^{(3)}$&$\left[-0.006,0.011\right]$&$\left[-0.006,0.011\right]$  &$\left[-0.117,0.027\right]$ \\[+0.05cm]
${\cal O}_{eu}$&$\left[-0.036,0.026\right]$&$\left[-0.046,0.024\right]$  &$\left[-0.044,0.117\right]$ \\[+0.05cm]
${\cal O}_{ed}$&$\left[-0.073,0.035\right]$&$\left[-0.074,0.037\right]$  &$\left[-0.128,0.086\right]$ \\[+0.05cm]
${\cal O}_{lu}$&$\left[-0.029,0.071\right]$&$\left[-0.035,0.075\right]$  &$\left[-0.053,0.095\right]$ \\[+0.05cm]
${\cal O}_{ld}$&$\left[-0.023,0.073\right]$&$\left[-0.021,0.083\right]$  &$\left[-0.117,0.094\right]$ \\[+0.05cm]
${\cal O}_{qe}$&$\left[-0.038,0.013\right]$&$\left[-0.043,0.012\right]$  &$\left[-0.051,0.070\right]$\\
\bottomrule
\end{tabular}
\caption{Combination of the different $95\%$ C.L. limits on  lepton-quark contact interactions.\label{EW_Comb_Lim}}
\end{table}

Given how stringent the global constraints are and 
the fact that LHC searches with $\sqrt{s}=8$ TeV are
already competitive with EWPD bounds, it is worth considering the
ability of the LHC to measure these operators in di-lepton
searches during its high-energy phase. Furthermore, in case a 
departure from the SM prediction is observed, it would be crucial to
try to understand the origin of such a departure. It is clear that any
di-lepton search at the LHC can be only sensitive to the combination
of operators described by the coefficients in
Eq.~(\ref{AB:coefficients}). Nevertheless, a very simple study of angular
distributions can discriminate between contributions that are mostly 
forward, mostly backward or symmetric. In order to test this, we have generated
di-muon events at $\sqrt{s}=14$ TeV and computed the observed number
of events and a forward-backward asymmetry, defined as
\begin{equation}
A_{FB}=\frac
{\sigma(\overline{\Delta \eta}>0)
 -\sigma(\overline{\Delta \eta}<0) }
{\sigma(\overline{\Delta \eta}>0)
 +\sigma(\overline{\Delta \eta}<0) },
\end{equation}
where $\overline{\Delta \eta}\equiv (\eta_{l^-}
- \eta_{l^+})/(\eta_{l^-} + \eta_{l^+})$ is positive (negative) for a
forward (backward) negatively charged lepton in the center of mass
frame, with respect to the direction of the incoming quark (when this
direction is
estimated by the beam axis in the
direction of the di-lepton  momentum  in the lab frame). 

\begin{figure}[h]
\centering
\includegraphics[scale=0.9]{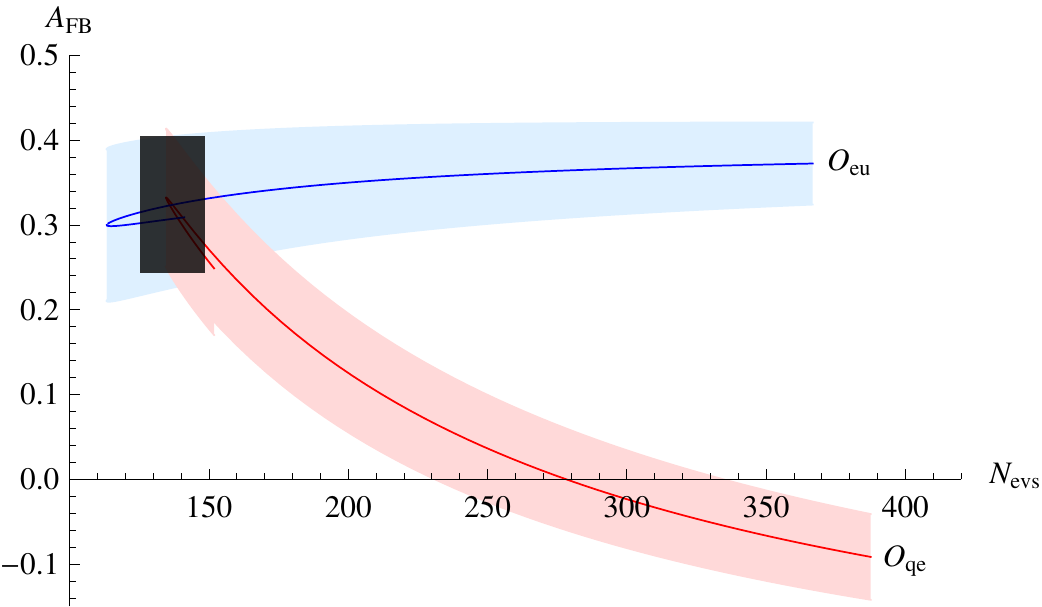}
\caption{\label{fig:AFBvsNev} 
Forward-backward asymmetry as a function
of the observed number of events for two representative operators.
We have considered $\sqrt{s}=14$ TeV with
$300$ fb$^{-1}$ of integrated luminosity and $M_{\ell^+\ell^-}\geq 1.8$
TeV. The coefficients of the
operators are varied within the current allowed values. The bands
represent the 1 $\sigma$ uncertainty on the asymmetry. The SM result
is represented, with 1 $\sigma$ uncertainties with a gray rectangle.}
\end{figure}

We show in Fig.~\ref{fig:AFBvsNev} the di-muon
$A_{FB}$ as a function of the observed number of events, both computed
with di-muon candidates with $M_{\mu^+ \mu^-}\geq 1.8$ TeV, for two
representantive operators, at the LHC with
$\sqrt{s}=14$ TeV and an integrated luminosity of 300 fb$^{-1}$. As a
reference we also plot the expected values for the SM. In the figure
we have varied the coefficients of the different operators within
their current limits. Both discovery in terms of number of events and
discrimination among different operators are clearly possible at this
center of mass energy.

In summary, we have provided a general parameterization of the effect
of an arbitrary number of lepton-quark contact interactions on
di-lepton production at hadron colliders. This is expressed in the
form of a master equation, 
Eqs.~(\ref{master:formula}) and (\ref{AB:coefficients}), in terms of a small
number of observable-dependent parameters (4 plus the SM prediction 
for forward-backward symmetric observables). 
Once these parameters have been computed for
the particular observable considered, the bounds on an arbitrary
combination of lepton-quark contact interactions can be
obtained. We have also found that it is important to consider more
than one bin in di-lepton invariant masses as different operators are
more efficiently constrained at different values of the di-lepton
invariant mass.  We have shown how to obtain such constraints by
combining LHC searches with indirect constraints from EWPD,
assuming one operator at a time. LHC searches are already
sometimes competitive and quite often complementary to EWPD. The very
stringent global constraints that we have obtained still leave room
for a discovery at the high-energy phase of the LHC. We have also
discussed how one can use angular observables to distinguish among different
classes of lepton-quark contact interactions.

\section*{Acknowledgments}
We thank F. del \'Aguila for useful comments. 
The work of J.B. has been supported in part by the U.S. National Science Foundation
under Grant PHY-1215979.
The work of M.C. and J.S. has been partially supported by MINECO projects
FPA2006-05294 and FPA2010-17915, by Junta de Andaluc\'{\i}a grants FQM
101 and FQM 6552. M.C. is also supported by the MINECO under the FPU program.


\end{document}